\documentclass[reprint,amsmath,amssymb,aps,showkeys,prb]{revtex4-2}

\usepackage{float}
\usepackage{caption}
\usepackage{subcaption}
\usepackage{graphicx}
\usepackage{dcolumn}
\usepackage{bm}
\usepackage{rotating}

\begin{document}
	
\title{"Adiabatic" Elastic Constants in Hubbard-Corrected Density-Functional Theory DFT+U: case UO$_2$}
\author{Mahmoud Payami }
\email{mpayami@aeoi.org.ir}
\author{Samira Sheykhi}
\affiliation{School of Physics \& Accelerators, Nuclear Science and Technology Research Institute, AEOI,\\ 
	P.~O.~Box~14395-836, Tehran, Iran}

\begin{abstract}
Since in DFT+U there are multiple self-consistent electronic solutions, the so called metastable states, the elastic constants computed from stress-vs-strain will be incorrect if some of the strained configurations fall into a different local electronic minimum than the equilibrium non-strained state. So, it is crucial to carefully take steps to keep the same electronic Hubbard occupation branch when computing the stresses for small strained geometries. In this work, we have explained this "adiabatic" method of calculation for elastic constants and applied for UO$_2$ crystal described within two different unit cells of cubic 12-atom and tetragonal 6-atom basis sets. The calculation results for the two different unit cells are the same within 0.1 GPa, and agreement with experiment is excellent.    
\end{abstract}

\keywords{DFT+U; Metastable states; Hubbard occupation branch; Stress; Strain; Elastic constant; Density-functional theory.}

\maketitle

\section{Introduction}\label{sec1}
\begin{figure}[ht]
	\centering
	\includegraphics[width=0.9\linewidth]{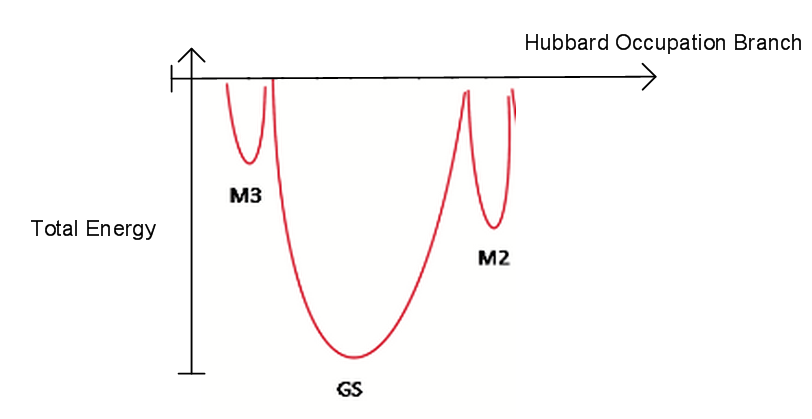}
	\caption{Schematic representation of the multi-valued energy functional with respect to different Hubbard occupation branches. The lowest lying minimum corresponds to the ground state and other local minima correspond to metastable states.}\label{fig1}
\end{figure}

In DFT+U \cite{dudarev1998,coco-degironc2005,payami-sheikhi23,payami24}, because of the added orbital-dependent terms to DFT functional \cite{hohenberg1964,kohn1965self}, the energy landscape will be multi-valued with respect to occupation matrices as shown schematically in Fig.~\ref{fig1}. 

The ground state, which is the lowest energy state out of many local minima, is determined via occupation matrix control (OMC) method \cite{payami-smcomc23}. In DFT+U, any small changes in ionic positions induced by strain or other dynamics, can flip the system to a different local minimum with a different occupancy pattern. Elastic constants are derivatives of stress with respect to strain, and if the underlying electronic state jumps discontinuously under strain, the calculated derivative is not the derivative in the same branch and will be incorrect. Keeping the occupation matrix fixed or consistently seeded and constrained, ensures that the differentiating is done in the same electronic branch with no discontinuity.
In section~\ref{sec2}, we explain mathematically the calculation method of adiabatic elastic constants. Section~\ref{sec3} is devoted to calculation details. In section~\ref{sec4} the results of application of the method to elastic constants of UO$_2$ crystal described within two different unit cells of cubic 12-atom and tetragonal 6-atom basis sets are presented; and finally section~\ref{sec5} concludes the work.      

\section{Adiabatic Elastic Constants}\label{sec2}
The adiabatic zero-temperature static elastic constants in general DFT (no Hubbard correction) are the second derivatives of the ground-state energy with respect to strain as:

\begin{equation}\label{eq1}
	C_{ijkl}=\left. \frac{1}{\Omega_0}\frac{\partial^2 E^{\text{DFT}}_{tot}}{\partial \varepsilon_{ij} \partial \varepsilon_{kl}}\right|_{\varepsilon=0}
\end{equation}
where $\Omega_0$ is the equilibrium cell volume and $\varepsilon_{ij}$ is the symmetric strain tensor. Equivalently, via the Hellmann-Feynman theorem,

\begin{equation}\label{eq2}
		C_{ijkl}=\left. \frac{\partial \sigma_{ij}}{\partial \varepsilon_{kl}}\right|_{\varepsilon=0},
\end{equation}
with $\sigma_{ij}$ the stress tensor.

Adding orbital-dependent Hubbard correction to DFT energy, the DFT+U total energy is given by 
\begin{equation}\label{eq3}
	E^{\text{DFT+U}}[n,\{n^I\}]=E^{\text{DFT}}[n]+E^{U}[\{n^I\}],
\end{equation}
where, $n(\bf r)$ is the total electron density, $n^I$ are the on-site occupation matrices for correlated Hubbard orbitals on atom $I$, and $E^U$ is the Hubbard correction (Dudarev type, Liechtenstein-type, etc.). For example, in Dudarev's simplified form we have \cite{dudarev1998,coco-degironc2005}:
\begin{equation}\label{eq4}
	E^U=\sum_{I,\sigma}\frac{U^I}{2} {\rm Tr[{\bf n}^{\it I\sigma}({\bf 1} - {\bf n}^{\it I\sigma}})].
\end{equation}

Thus the stress tensor acquires an additional Hubbard contribution:
\begin{equation}\label{eq5}
	\sigma_{ij}^{\text{DFT+U}} = \sigma_{ij}^{\text{DFT}} + \sigma_{ij}^{U},
\end{equation}
where 
\begin{equation}\label{eq6}
	\sigma_{ij}^U = \frac{1}{\Omega} \frac{\partial E^U}{\partial \varepsilon_{ij}}.
\end{equation}

The Hubbard occupation matrices $\{n^I\}$ play important role in the derivatives.
They are variational degrees of freedom in DFT+U such that for each geometry consisting of "lattice+ions", the self-consistent solution finds a stationary point on both $n(\bf r)$ and $\{n^I\}$.
Formally, the elastic constants should be computed from the second derivative of the minimum energy branch:

\begin{equation}\label{eq7}
	C_{ijkl}=\left. \frac{1}{\Omega_0} \frac{\partial^2}{\partial \varepsilon_{ij} \partial \varepsilon_{kl}} \left[ \min_{\{n^I\}} E^{\text{DFT+U}}[n,\{n^I\}]\right] \right|_{\varepsilon=0}.
\end{equation} 

In DFT+U, the minimization over $\{n^I\}$ has multiple local minima corresponding to metastable occupation matrices. 
Then, derivatives are well defined only if one stays on the same branch of $\{n^I\}$. Mathematically, one must treat $\{n^I\}$ as implicit functions of strain constrained to the chosen minimum branch:

\begin{equation}\label{eq8}
	n^I(\varepsilon) \;\;\;\; \text{with} \;\;\; \left. \frac{\partial E^{\text{DFT+U}}}{\partial n^I} \right|_{n^I(\varepsilon)} = 0.
\end{equation}
Thus, 
\begin{equation}\label{eq9}
	C_{ijkl} = \frac{1}{\Omega_0} \left[ \frac{\partial^2 E^{\text{DFT+U}}}{\partial \varepsilon_{ij} \partial \varepsilon_{kl}} + \sum_I \frac{\partial^2 E^{\text{DFT+U}}}{\partial\varepsilon_{ij}\partial n^I} \frac{\partial n^I}{\partial \varepsilon_{kl}} \right]_{\varepsilon=0,\; n^I=n^I_0}.
\end{equation}

Under the condition of "fixed-occupation branch", at self-consistency, the first-order derivatives with respect to occupations vanish:
\begin{equation}\label{eq10}
	\left. \frac{\partial E^{\text{DFT+U}}}{\partial n^I} \right|_{n^I_0} = 0.
\end{equation}
Therefore, by implicit function theorem, the term involving $\partial n^I/\partial\varepsilon$ drops out if one stays on the same local minimum branch.
So, in practice, one just needs to ensure the occupations do not "jump", and then the usual stress-strain or energy-strain formula is valid.

To sum up: {\it i}) The occupation matrices act as auxiliary variational parameters; {\it ii}) If one stays on the same minimum branch, their implicit strain-dependence does not contribute to the second derivative (elastic constants); and {\it iii}) If the system hops to another local minimum, the derivative becomes discontinuous, and the computed "elastic constants" will be unphysical.

So, the theoretical formulation for the "correct workflow" is:
\begin{equation}\label{eq11}
	C_{ijkl} = \frac{1}{\Omega_0} \left. \frac{\partial^2 E^{\text{DFT+U}}}{\partial \varepsilon_{ij} \partial \varepsilon_{kl}} \right|_{\varepsilon=0,\; \{n^I\}=\{n^I_0\}},
\end{equation} 
where $\{n^I_0\}$ are the occupation matrices of the equilibrium ground state.

\section{Calculation Details}\label{sec3}
\begin{figure}[ht]
	\centering
	\includegraphics[width=0.9\linewidth]{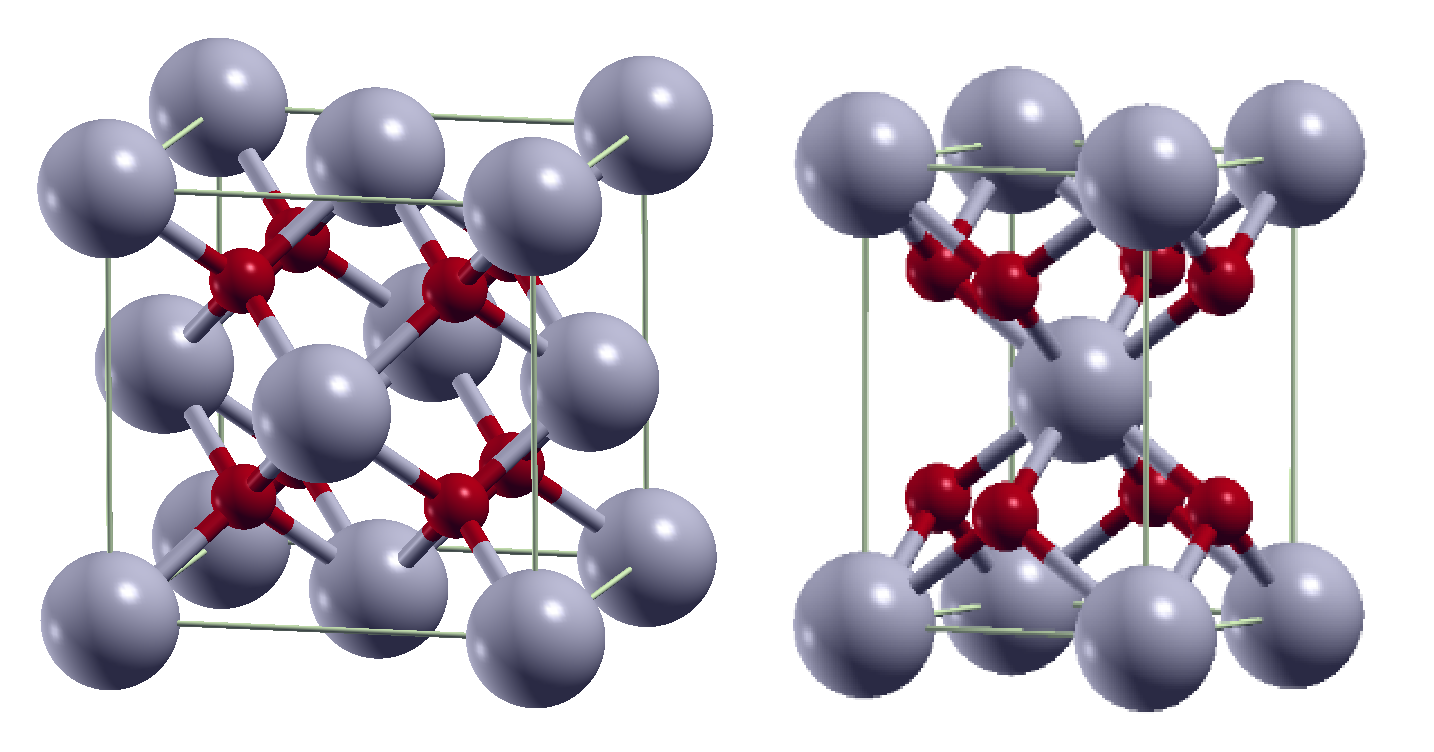}
	\caption{UO$_2$ crystal structure with cubic 12-atom unit cell (left) and with simple tetragonal 6-atom unit cell (right). Large grey and small red balls represent uranium and oxygen atoms, respectively.}\label{fig2_new}
\end{figure}

The uranium dioxide crystal can be described with good accuracy using the cubic $Fm\bar{3}m$ space group (space group number 225) with a lattice constant $a = 5.47\AA$, as shown in the left of Fig.~\ref{fig2_new}, or by a simple tetragonal unit cell with a 6-atom basis set shown in the right of Fig.~\ref{fig2_new}. The antiferromagnetic (AFM) structure of uranium atoms is modeled by the simple one-dimensional AFM in $z$-direction.  

The sets of lattice vectors $\{{\bf a_1}, {\bf a_2}, {\bf a_3}\}$ for cubic and $\{{\bf a^\prime_1}, {\bf a^\prime_2}, {\bf a^\prime_3}\}$ for tetragonal unit cells are shown in Fig.~\ref{fig3_new} and are defined in Cartesian coordinates as: 

\begin{figure}[ht]
	\centering
	\includegraphics[width=0.7\linewidth]{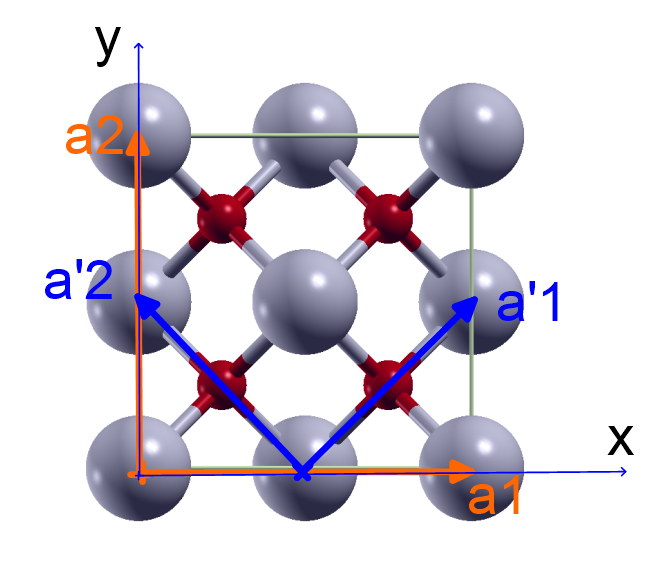}
	\caption{Top view of cubic 12-atom and simple tetragonal 6-atom unit cells. Lattice vectors $\{{\bf a_1}, {\bf a_2}, {\bf a_3}\}$ and $\{{\bf a^\prime_1}, {\bf a^\prime_2}, {\bf a^\prime_3}\}$ describe the cubic and tetragonal cells, respectively.}\label{fig3_new}
\end{figure}

\begin{equation}\label{eq12}
	{\bf a_1} = a(1,0,0) \;\;\; {\bf a_2} = a(0,1,0) \;\;\; {\bf a_3} = a(0,0,1)
\end{equation}

\begin{equation}\label{eq13}
	{\bf a^\prime_1} = \frac{a}{2}(1,1,0) \;\;\; {\bf a^\prime_2} = \frac{a}{2}(-1,1,0) \;\;\; {\bf a^\prime_3} = a(0,0,1)
\end{equation}

In this work, using DFT+U, the elastic constants of UO$_2$ is calculated using both cubic and tetragonal unit cells. The results are in excellent agreement with each other and with experiment.

\vspace{5pt}
{\noindent \bf DFT+U Calculations}\\
For the electron-ion interactions we have used scalar-relativistic ultra-soft pseudo-potentials (USPP) with PBEsol approximation for the XC, generated using "atomic" code. The valence configurations U($6s^2,\, 6p^6,\, 7s^2,\, 7p^0,\, 6d^1,\, 5f^3 $) and O($2s^2,\, 2p^4 $) were used in the USPP generation. All DFT+U calculations were based on the solution of the KS equations using the Quantum-ESPRESSO (QE) code package \cite{qe-2009,qe-2020}.  
 Kinetic energy cutoffs for the plane-wave expansions
were chosen as 90 and 720~Ry for the wave-functions and densities, respectively. The smearing method of Marzari-Vanderbilt for the occupations with a width of 0.001~Ry were used. 
For the unstrained and strained cell Brillouin-zone integrations, $6\times 6\times 6$ and $10\times 10\times 7$ grids were used for cubic and tetragonal cells, respectively;  All geometries were fully optimized for total residual pressures on unit cells to within 0.5 kbar, and residual forces on atoms to within 10$^{-3}$~mRy/a.u. For Hubbard orbitals, we have used the "atomic" projection operators for the expansion of KS orbitals. The value of Hubbard parameter for the $5f$ orbital of uranium atoms is set to 2.45 eV \cite{payami23} which gives the lattice constant of optimized structure equal to the experimental value 5.47\AA.

\vspace{5pt}
{\noindent \bf Elastic Constants Calculations}\\
To calculate the elastic constants, first the geometries were fully optimized so that the forces on atoms and stress on crystal lattice vanished to a good accuracy. Then, we apply specific changes on the lattice vectors to make system strained and let the atoms relax to their new equilibrium positions. Afterwards, for the strained system the stress tensor is calculated. Finally, having the stress and strain tensors at hand, the elastic constants were calculated via the stress-strain formula. The second-rank stress and strain tensors, which are denoted respectively by $\sigma$ and $\varepsilon$, are related with a fourth-rank tensor as\cite{nye85}:

\begin{equation}\label{eq14}
	\sigma_{ij}=\sum_{kl}C_{ijkl}\varepsilon_{kl},
\end{equation}   
in which $i,j,k,l=x,y,z$ are Cartesian coordinates.
Both stress and strain tensors are symmetric and in the Voigt notation, under the mapping of indices $xx\mapsto 1 , yy\mapsto 2,zz\mapsto 3, yz\mapsto 4, xz\mapsto 5, xy\mapsto 6$, the above relation can be represented in matrix form as:

\begin{equation}\label{eq15}
	\begin{bmatrix}
		\sigma_1\\
		\sigma_2\\
		\sigma_3\\
		\sigma_4\\
		\sigma_5\\
		\sigma_6 
	\end{bmatrix}
	=
	\left[
	\begin{array}{cccccc}
		C_{11} & C_{12} & C_{13} & C_{14} & C_{15} & C_{16} \\
		C_{21} & C_{22} & C_{23} & C_{24} & C_{25} & C_{26} \\
		C_{31} & C_{32} & C_{33} & C_{34} & C_{35} & C_{36} \\
		C_{41} & C_{42} & C_{43} & C_{44} & C_{45} & C_{46} \\
		C_{51} & C_{52} & C_{53} & C_{54} & C_{55} & C_{56} \\
		C_{61} & C_{62} & C_{63} & C_{64} & C_{65} & C_{66} \\
	\end{array}
	\right]
	\begin{bmatrix}
		\varepsilon_1\\
		\varepsilon_2\\
		\varepsilon_3\\
		2\varepsilon_4\\
		2\varepsilon_5\\
		2\varepsilon_6\\
	\end{bmatrix}
\end{equation}  

The elastic tensor in this notation is a $6\times 6$ symmetric matrix with 21 independent components. If we denote the lattice vectors of optimized geometry by $\{{\bf a_1}, {\bf a_2}, {\bf a_3}\}$, then applying the following 6 independent transformation $F_i$ (one at a time) on optimized lattice vectors, we will have 6 strained states:

{\small
\begin{widetext}
	\begin{equation*}\label{eq16}\tag{16}
		\begin{split}
			F_1=\left[\begin{array}{ccc}
				1+\delta & 0 & 0 \\
				0 & 1 & 0 \\
				0 & 0 & 1 \\
			\end{array}\right]
			,
			F_2=\left[\begin{array}{ccc}
				1 & 0 & 0 \\
				0 & 1+\delta & 0 \\
				0 & 0 & 1 \\
			\end{array}\right]
			,
			F_3=\left[\begin{array}{ccc}
				1 & 0 & 0 \\
				0 & 1 & 0 \\
				0 & 0 & 1+\delta \\
			\end{array}\right]
		,
			F_4=\left[\begin{array}{ccc}
				1 & 0 & 0 \\
				0 & 1 & \delta \\
				0 & 0 & 1 \\
			\end{array}\right]
			,
			F_5=\left[\begin{array}{ccc}
				1 & 0 & \delta \\
				0 & 1 & 0 \\
				0 & 0 & 1 \\
			\end{array}\right]
			,
			F_6=\left[\begin{array}{ccc}
				1 & \delta & 0 \\
				0 & 1 & 0 \\
				0 & 0 & 1 \\
			\end{array}\right] \;\;\;\;\;\;\;\;\;\;\;\;\;\;\;\;\;\;
		\end{split}
	\end{equation*}
\end{widetext} 
}
For each of the 6 strained states, we calculate the stress tensors for 4 different $\delta$ values, $\delta\in\{-0.01, -0.005, +0.005, +0.10 \}$, and then fit a straight line to extract the elastic constant from the linear coefficient.  

\section{Results and discussions}\label{sec4}
The geometries of cubic and tetragonal cells are fully optimized so that the forces on atoms and stress components on lattices vanish within a good accuracy.    

To calculate the elastic constants, we applied different strains and calculated the corresponding stresses. Applying the strain operator $F_1$, we extract the 6 elastic constants: $C_{11}, C_{21}, C_{31}, C_{41}, C_{51}, C_{61} $.  
In addition, applying $F_2$, we obtain 6 other elastic constants: $C_{12}, C_{22}, C_{32}, C_{42}, C_{52}, C_{62} $, and so forth. One may benefit the symmetry relations to reduce the calculations, but we did not.
In Eq.~(\ref{eq17}) and Eq.(~\ref{eq18}), we have presented the results for cubic and tetragonal cells, respectively and in cubic results we have mentioned the experimental values \cite{wachtman65} in parentheses. Comparison shows very good agreement between the cubic and tetragonal results and also experiment. 
\begin{widetext}
\begin{equation*}\label{eq17}\tag{17}
{\bf C_{\text{cubic}}}	=
	\left[
	\begin{array}{cccccc}
		371.879 (395\pm 1) & 121.047 (121\pm 2) & 124.913 & 0.0 & 0.0 & 0.0  \\ 
        121.068 & 371.856 & 124.913 & 0.0 & 0.0 & 0.0  \\ 
        124.869 & 124.851 & 377.602 & 0.0 & 0.0 & 0.0  \\ 
        0.0 & 0.0 & 0.0 & 72.491 (64.1\pm 1) & 0.0 & 0.0  \\ 
        0.0 & 0.0 & 0.0 & 0.0 & 72.492 & 0.0  \\ 
        0.0 & 0.0 & 0.0 & 0.0 & 0.0 & 70.642  \\  
	\end{array}
	\right]
\end{equation*}  
\end{widetext}

\begin{widetext}
	\begin{equation*}\label{eq18}\tag{18}
		{\bf C_{\text{tetragonal}}}	=
		\left[
		\begin{array}{cccccc}
			371.906 & 121.095 & 124.901 & 0.0 & 0.0 & 0.0  \\ 
			121.103 & 371.897 & 124.901 & 0.0 & 0.0 & 0.0  \\ 
			124.890 & 124.890 & 377.564 & 0.0 & 0.0 & 0.0  \\ 
			0.0   & 0.0   & 0.0   & 72.485& 0.0 & 0.0  \\ 
			0.0   & 0.0   & 0.0   & 0.0 & 72.485& 0.0  \\ 
			0.0   & 0.0   & 0.0   & 0.0 & 0.0 & 70.640  \\  
		\end{array}
		\right]
	\end{equation*}  
\end{widetext}
The elastic constants of Eq.~(\ref{eq17}) and Eq.~(\ref{eq18}) show excellent agreements.
Detailed discussions on the cubic results was presented in Ref.~\cite{PSB-25}. 

\section{Conclusions}\label{sec5}
Because of multiple self-consistent electronic solutions in DFT+U, the so called metastable states, the elastic constants obtained from stress-vs-strain will be wrong if some of the strained configurations fall into a different local electronic minimum than the equilibrium non-strained state. Then the derivatives of total energies with respect to strain will be discontinuous. Therefore, one should stay at the same electronic Hubbard occupation branch when computing the stresses for small strained geometries. This so-called "adiabatic" elastic constants were calculated for UO$_2$ crystal described with two different cubic and tetragonal unit cells, and the results showed excellent agreement with each other and with experiment.

\section*{Acknowledgement}
This work is part of research program in School of Physics and Accelerators, NSTRI, AEOI.  

\section*{Data availability }
The raw or processed data required to reproduce these results can be shared with anybody interested upon 
sending an email to M. Payami.


\vspace*{2cm}
\section*{References}

\begin{thebibliography}{0}
\bibitem{dudarev1998} S. L. Dudarev, G. A. Botton, S. Y. Savrasov, C. J. Humphreys, and A. P. Sutton, {\it Phys. Rev.} B{\bf 57}, 1505 (1998).

\bibitem{coco-degironc2005} M. Cococcioni and S. de Gironcoli, {\it Phys. Rev.} B {\bf 71}, 035105 (2005).
	
\bibitem{payami-sheikhi23} M. Payami, S. Sheykhi, and M.R. Basaadat, https://doi.org/10.48550/arXiv.2306.06266 (2023), and references therein.

\bibitem{payami24}  M. Payami, https://doi.org/10.48550/arXiv.2401.00864 (2024), and references therein.

\bibitem{hohenberg1964} P. Hohenberg and W. Kohn, {\it Phys. Rev.} {\bf 136}, B864 (1964).

\bibitem{kohn1965self} W. Kohn and L. J. Sham, {\it Phys. Rev.} {\bf 140}, A1133 (1965).

\bibitem{payami-smcomc23} M. Payami, https://doi.org/10.47176/ijpr.23.3.11820, and references therein.









\bibitem{qe-2009} P. Giannozzi, S. Baroni, N. Bonini, et. al., {\it J. Phys.: Condensed Matt.} {\bf 21}, 395502 (2009).

\bibitem{qe-2020} P. Giannozzi, O. Baseggio, P. Bonfà, et. al., {\it J. Chem. Phys.} {\bf 152}, 154105 (2020).

\bibitem{payami23}  M. Payami, https://doi.org/10.48550/arXiv.2302.13381 (2023), and references therein.

\bibitem{nye85} J. F. Nye, {\it Physical Properties of Solids- Their Representation by Tensors and
Matrices}, OUP (1985).

\bibitem{wachtman65} J. B. Wachtman Jr., M. L. Wheat, H. J. Anderson, J. L. Bates, {\it J. Nucl. Mat.}
	{\bf 16} (1), 39-41 (1965).

\bibitem{PSB-25} M. Payami, S. Sheykhi, and M. R. Basaadat, https://doi.org/10.48550/arXiv.2504.08468.

\end{thebibliography}
\end{document}